\documentclass[12pt,preprint]{aastex}

\usepackage{epsfig}
\input psfig

\shorttitle{The Mt~John $\alpha$ Cen Program}
\shortauthors{Endl et al.}

\begin{document}

\title{The Mt John University Observatory Search For Earth-mass Planets In The Habitable Zone Of $\alpha$ Centauri}  

\author{Michael Endl}
\affil{McDonald Observatory, The University of Texas at Austin,
    Austin, TX 78712}
\email{mike@astro.as.utexas.edu}
\author{Christoph Bergmann}
\affil{Department of Physics \& Astronomy, The University of Canterbury,
    Christchurch 8041, New Zealand}
\author{John Hearnshaw}
\affil{Department of Physics \& Astronomy, The University of Canterbury,
    Christchurch 8041, New Zealand}
\author{Stuart I. Barnes}
\affil{McDonald Observatory, The University of Texas at Austin,
    Austin, TX 78712}
\affil{Department of Physics \& Astronomy, The University of Canterbury,
    Christchurch 8041, New Zealand}
\author{Robert A. Wittenmyer}
\affil{Department of Astrophysics and Optics, School of Physics, University of New South Wales, Sydney, Australia}
\author{David Ramm}
\affil{Department of Physics \& Astronomy, The University of Canterbury,
    Christchurch 8041, New Zealand}
\author{Pam Kilmartin}
\affil{Department of Physics \& Astronomy, The University of Canterbury,
    Christchurch 8041, New Zealand}
\author{Fraser Gunn}
\affil{Department of Physics \& Astronomy, The University of Canterbury,
    Christchurch 8041, New Zealand}
\author{Erik Brogt}
\affil{Academic Development Group, The University of Canterbury,
    Christchurch 8041, New Zealand}

\begin{abstract}
The ``holy grail'' in planet hunting is the detection of an Earth-analog: a planet with similar mass as the
Earth and an orbit inside the habitable zone. If we can find such an Earth-analog around one of the stars
in the immediate solar neighborhood, we could potentially even study it in such great detail to address the
question of its potential habitability. 
Several groups have focused their planet detection efforts on the nearest stars. Our team is currently 
performing an intensive observing campaign on the $\alpha$ Centauri system using the {\sc Hercules} spectrograph
at the 1-m McLellan telescope at Mt John University Observatory (MJUO) in New Zealand. 
The goal of our project is to obtain such a large number of radial velocity measurements with sufficiently 
high temporal sampling to become sensitive to signals of Earth-mass planets in the habitable zones of the
two stars in this binary system.
Over the past years, we have collected more than 45,000 spectra for both stars combined. These data
are currently processed by an advanced version of our radial velocity reduction pipeline, which eliminates the 
effect of spectral cross-contamination. Here we present simulations of the expected detection sensitivity to 
low-mass planets in the habitable zone by the {\sc Hercules} program for various noise levels. We also discuss
our expected sensitivity to the purported Earth-mass planet in an 3.24-d orbit announced by Dumusque et al.~(2012).
\end{abstract}

\keywords{planetary system --- stars: individual ($\alpha$~Cen~A, $\alpha$~Cen~B) --- techniques: radial
velocities}

\section{Introduction}

The search for a true Earth-analog planet is one of the boldest scientific and intellectual 
endeavors ever undertaken by humankind. If such a planet can be found orbiting a nearby Sun-like star, it
will constitute an ideal target for extensive follow-up studies from the 
ground and with future space missions. These follow-up studies can include a detailed characterization 
of the planetary system and, ultimately, a search for bio-signatures in the atmosphere of an Earth-like planet 
in the habitable zone. For the far future, many decades to centuries from now, one can even imagine that the first 
interstellar probe will be launched to travel to one of the 
systems, where we found evidence for a nearby Earth twin. The discovery of such a planet will have
an unprecedented cultural as well as scientific impact.

NASA's {\it Kepler} mission (Borucki et al.~2010) has been extremely successful in finding small, possibly rocky planets
orbiting stars in the {\it Kepler} search field. Some of them even reside within the circumstellar habitable zones: e.g.
Kepler-22b (Borucki et al.~2012) and two planets in the Kepler-62 system (Borucki et al.~2013). One of the most
significant result from {\it Kepler} is the planet occurrence rates, which show that small radius planets are 
quite frequent and outnumber the giant planets to a large extent
(e.g. Howard et al.~2012, Fressin et al.~2013, Dressing et al.~2013, Petigura et al.~2013). We are
tempted to extrapolate from these high {\it Kepler} planet frequencies to the immediate solar neighborhood and conclude
that many nearby stars, possibly even the nearest star to the Sun, are orbited by one or more Earth-like planets.      

The precision of stellar radial velocity (RV) measurements has steadily improved from a modest 15\,m\,s$^{-1}$ (Campbell \& Walker~1979), more than 
three decades ago, to a routine 3\,m\,s$^{-1}$ (Butler et al.~1996), and in the best case, with the highly stabilized
HARPS spectrograph (Mayor et al.~2003), even 1\,m\,s$^{-1}$ or better. The
discovery space of the RV method was, therefore, extended from the giant planet domain down
to Neptunes and super-Earths (with minimum masses between 2 and 10~M$_{\oplus}$). 
In terms of RV precision, we are still more than an order of magnitude from the 0.09\,m\,s$^{-1}$ RV amplitude of an  
Earth at 1~AU orbiting a G-type star. Future projects aim for RV precision of $0.1$\,m\,s$^{-1}$, but they are still years away
from being operational. ESPRESSO (Pepe et al.~2014) is currently under construction for the ESO Very Large Telescope
and G-CLEF (Szentgyorgyi et al.~2012) has been selected as a first-light instrument for the Giant Magellan Telescope (GMT). 

However, there is an alternative to extreme precision:
with a large enough number of measurements, even signals with amplitudes orders of magnitude
below the individual measurement uncertainties can be detected with high significance. Instead of waiting for
the new instruments to be deployed, several groups have started ambitious RV programs that observe a small
sample of suitable stars in the solar neighborhood with high temporal cadence. 
Owing to the extreme observational effort, these searches have to be dedicated to a few systems rather than to include
as many targets as possible. The need of RV searches to focus on nearby bright stars has an attractive side-effect:
the targets are all very close to the Sun, unlike the {\it Kepler} targets or microlensing systems, which are at typical 
distances of several hundreds or thousands of parsecs.  
The HARPS team has focused on ten solar-type
stars and reported very low-mass planets around HD~20794, HD~85512 and HD~192310 (Pepe et al.~2011). 
In order to properly perform such an
RV search for low-mass planets, it is important 
to pay careful attention to RV signals that are {\it intrinsic} to the star and that can have a larger amplitude than a
planetary signal.   
   
Special attention has been given to our closest neighbor in space, $\alpha$~Centauri, and several groups (see Table\,\ref{groups}) have
chosen this star system as the prime target of their planet detection efforts. Recently, Dumusque et al.~(2012) presented the case for
the presence of a low-mass planet in a 3.2~d orbit around $\alpha$~Cen~B using 459 highly precise RV measurements with HARPS obtained over 
a time span of 4 years. 
And, Tuomi et al.~(2012) discussed the possible existence of a 5-planet system around $\tau$~Ceti, another very close Sun-like star for
which no planets have yet been reported. These important results need to be confirmed by independent data and analysis. 
Indeed, Hatzes~(2013) re-analysed the HARPS RV results for $\alpha$~Cen~B using a different approach to filter out the stellar activity 
signals than that of Dumusque et al. and cast serious doubt on the reality, or at least on the planetary nature of the 3.2~d signal.  
In this paper we describe our $\alpha$~Cen program with the {\scshape Hercules} spectrograph at the McLellan 1\,m telescope at 
Mt~John University Observatory (MJUO) in New Zealand. 

\begin{deluxetable}{lll}
\tablecolumns{3}
\tablewidth{0pt}
\tablecaption{Precise radial velocity surveys that target the $\alpha$\,Cen system.
\label{groups}}
\tablehead{
\colhead{Site} & {Spectrograph/Telescope} & {Reference/Project website}
}
\startdata
La Silla & HARPS / ESO 3.6\,m & Pepe et al.(2011), Dumusque et al.~(2012)\\
CTIO & CHIRON / SMARTS 1.5\,m  & {\tiny http://exoplanets.astro.yale.edu/instrumentation/chiron.php} \\
MJUO & HERCULES / 1\,m McLellan & {\tiny http://www2.phys.canterbury.ac.nz/$\sim$physacp/index.html}\\
\hline
\enddata
\end{deluxetable}

\section{The $\alpha$ Centauri System}

The $\alpha$~Cen system is our closest neighbor in space ($d=1.347$~pc) and thus constitutes a target of fundamental importance
for any exoplanet search program. It is so close that a future spacecraft traveling at 0.1\,c reaches the system within 50 years.
The $\alpha$~Centauri binary consists of a G2V primary (HR 5459, HD 128620, $V=-0.01$) and a K1V secondary (HR~5460, HD~128621, $V=1.33$)
moving in an eccentric ($e = 0.518$) orbit with a semi-major axis of $a = 23$~AU and a period of almost 80 years (Heintz 1982, Pourbaix et al.~1999).
The possible third member of the system, Proxima Cen (M5V, $V=11.05$), is located at a much larger separation of $\approx 12000$~AU. 
It is not clear yet whether Proxima is indeed gravitationally bound to the inner binary (Wertheimer \& Laughlin 2006). 
Stringent upper mass limits for planets in the habitable zone of Proxima Cen have been presented by Endl \& K\"urster~(2008) and 
Zechmeister, K\"urster \& Endl~(2009).

The star $\alpha$~Cen~A is also very similar to our Sun. 
A mass ratio of the binary ($m_{\rm B}/m_{\rm A}=0.75\pm0.09$) was first presented by Murdoch \& Hearnshaw~(1993).
A recent study of the atmospheric parameters and abundances of the two stars was presented by Porto de Mello et al.~(2008). 
Both stars are almost twice as metal-rich as the Sun (England~1980, Furenlid \& Meylan~1990 and Porto de Mello et al.~2008).
The age of the system has been estimated to be between 5.6 and 6.5 Gyr (e.g. Eggenberger et al.~2004). 
The masses of the two stars have been determined with high accuracy by Pourbaix et al.~(2002): $M_{\rm A}$\,=\,1.105$\pm$0.007~$M_{\odot}$ and
$M_{\rm B}$\,=\,0.934$\pm$0.006~$M_{\odot}$. 

The $\alpha$~Cen system was included in the original 1992 target sample of the RV planet search at the ESO Coud\'e Echelle 
Spectrometer ({\scshape Ces}) (K\"urster et al.~1994). In Endl et al.~(2001) we presented our {\scshape Ces} RV data and a full analysis 
of the system including mass upper limits for planets. 
Based on the first five and a half years of {\scshape Ces} RV measurements, we were able to set stringent constraints on the presence of any gas giant 
planets in the
systems. We basically excluded the presence of any giant planet with a minimum mass greater than 3.5~$M_{\rm Jup}$. 
Moreover, we combined the RV-derived mass limits with the limits on stable orbits around each star imposed by the
presence of the other star from Wiegert \& Holman~(1997). They found that stable regions for planets orbiting in the binary plane (corresponding 
to a viewing angle $i\,=\,79.23^{o}$) extend to roughly 3~AU around each star, for prograde orbits, and to 4~AU, for retrograde orbits.
If the planets orbit in the binary plane, the RV measured $m \sin i$ values are close to the actual mass.  

The formation of terrestrial planets in the $\alpha$~Cen system was studied by Quintana et al.~(2002, 2007), 
Barbieri et al.~(2002), Quintana \& Lissauer~(2006), and
by Guedes et al.~(2008). These studies suggest that the formation of several Earth-mass planets at separations less than 2~AU 
was possible, despite the perturbing influence of the stellar companion. 

Guedes et al.\,simulated the RV detectability of these planets using synthetic RV data sets. They demonstrated that a 1.7~$M_{\oplus}$ planet 
with a period of 1.2 years can be clearly detected after 3 years of observations, if the noise distribution is sufficiently close to gaussian 
(see Fig.~5 in their paper). They used a noise level of 3\,m\,s$^{-1}$ and nearly 10$^{5}$ synthetic RV points.    

Thebault et al.~(2008, 2009), however, argue that these Earth-like planets might not have formed around $\alpha$~Cen~B as the mutual 
velocities of planetesimals were too high to allow accretion and continued growth to proto-planets beyond the planetesimal phase. These authors also state that the situation
would have been more benign for the formation of terrestrial planets if the binary was initially wider (by about 15~AU) or had a lower eccentricity than 
currently. On the other hand, Xie et al.~(2010) conclude that the formation of terrestrial planets around $\alpha$~Cen~B was possible, even in 
the current binary configuration. 

\section{The Mt~John University Observatory $\alpha$~Centauri Program}

The University of Canterbury in Christchurch, New Zealand, owns and operates the Mt~John University Observatory (MJUO) at the 
northern end of the Mackenzie Basin in the South Island. We selected MJUO as the site for this project, as 
$\alpha$~Cen, at lower culmination, is still 15 degrees above the horizon.
While other observatories are limited to observe $\alpha$~Cen for 9 to 10 months, we are able to obtain RV data for 12 
months a year. 
We are in a unique position that allows us optimal sampling of periods close to one year.

\subsection{Observing Strategy}

Terrestrial planets orbiting inside the classical habitable zone around F, G and K-type stars
only induce reflex velocities of $\approx$ 0.1\,m\,s$^{-1}$ on their host stars. This is still
well below the current state-of-the-art level of measurement precision.

But the sensitivity to low amplitude signals is not only a function of measurement precision, but also of
data quantity. With a large enough number of measurements, even signals with amplitudes below
the individual measurement uncertainties can be detected with high significance. Cochran \& Hatzes~(1996)
presented analytical expressions for the detection sensitivity of an RV survey of a given precision.
Their analytical relationship between noise and the total number of independent measurements ($N$) demonstrates that
at least $N$\,=\,100 is necessary to detect a signal with an amplitude equal to the noise.
If the errors are
sufficiently close to white noise, this trend continues toward lower amplitudes and with the appropriate
high value of $N$ even signals with amplitudes much lower than the noise can be detected. From this we see that,
to find Earth-mass planets with minimum masses of 0.5-2~$M_{\oplus}$ with habitable zone periods, $N$ has to be of the order of several 10$^{3}$ to 10$^{4}$,
depending on the measurement uncertainties.
This is the strategy that we adopted for the Mt~John program: to obtain a very large number of RV measurements at
a modest 2-3~m\,s$^{-1}$ precision level.

\begin{figure}[t]
\includegraphics[angle=0,scale=.45]{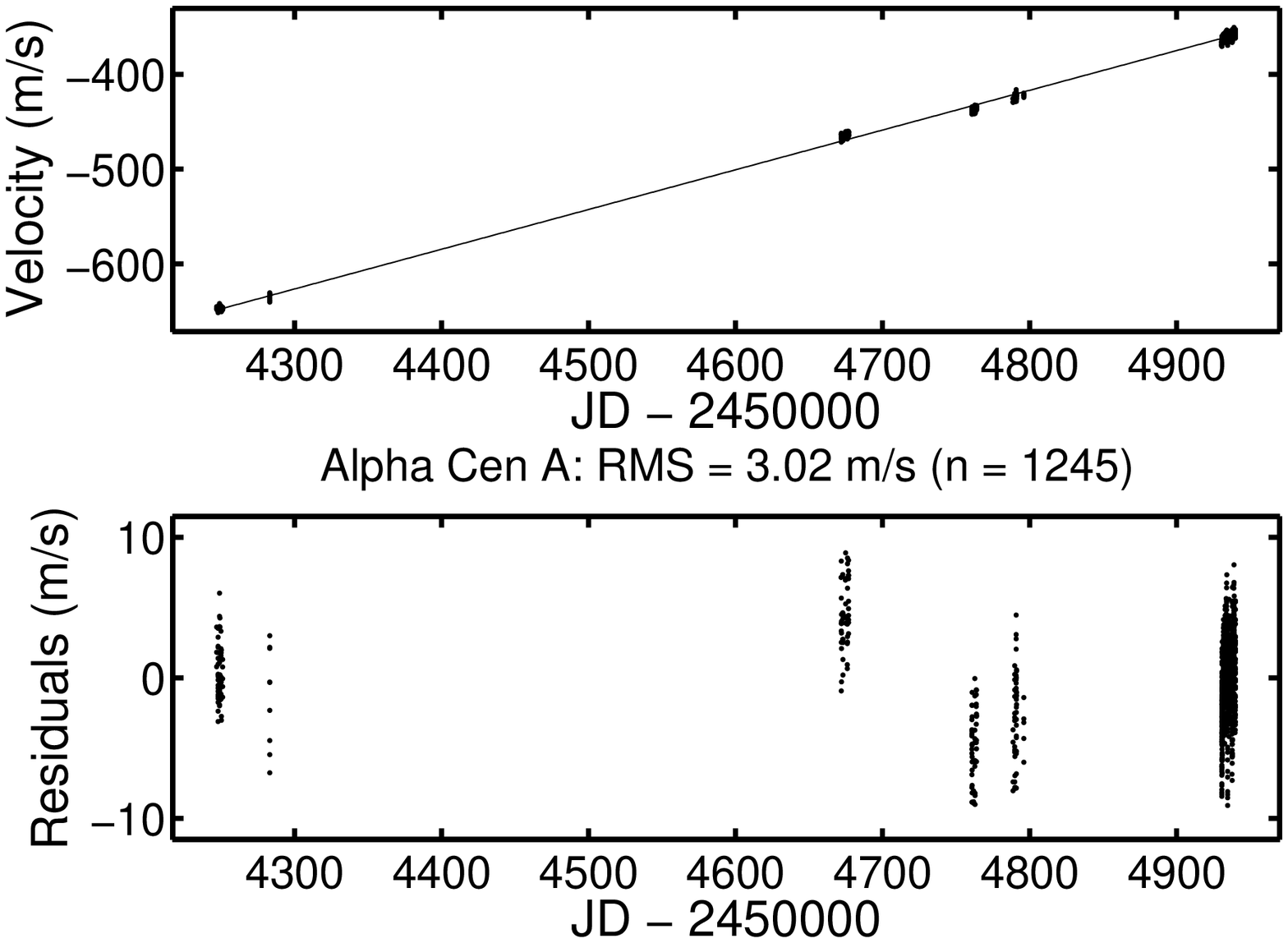}
\includegraphics[angle=0,scale=.45]{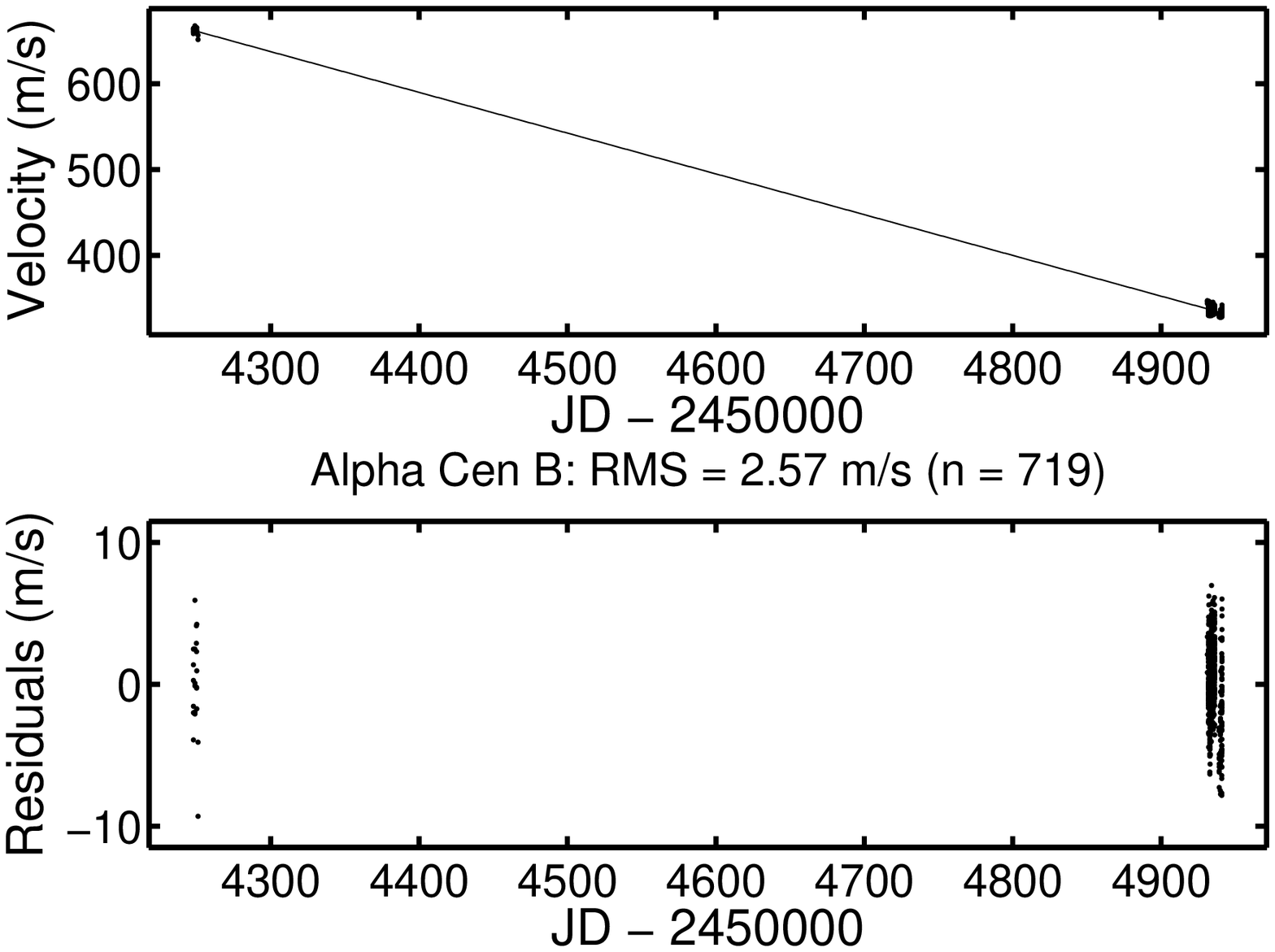}
\caption{
The 2007-2009 {\scshape Hercules}+I$_2$ RV data for $\alpha$~Cen~A \& B.
The top panels show the data with the strong slope due to the binary orbit (solid line) and the lower
panel displays the residuals. The residual scatter after removal of the slope is
2.5\,-\,3.0\,m\,s$^{-1}$. The data also show a larger level of RV variability for
A than for B (which is expected from their spectral types and level of magnetic activity).
\label{alpcen}}
\end{figure}

\begin{figure}
\parbox{10cm}
{\psfig{figure=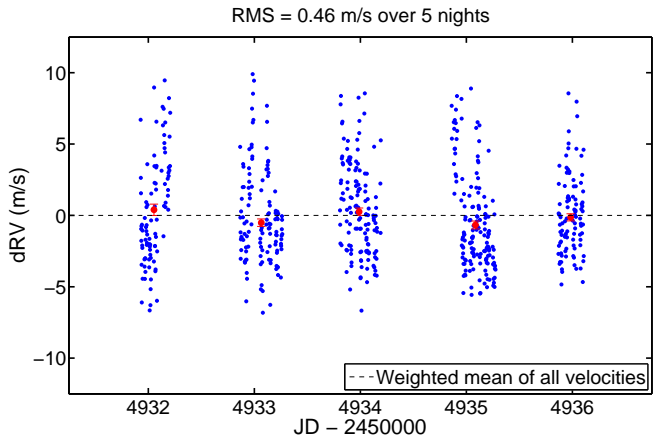, width=10cm, angle=0}}
\parbox{6cm}
{\caption{\small Five consecutive nights of
{\scshape Hercules} RV measurements for $\alpha$~Cen~B from our 2009 April run. The
$>500$ individual measurements have a total scatter of 2.71\,m\,s$^{-1}$ and show very little night to night
variations. The weighted mean values of these five nights have a standard deviation of only 0.5\,m\,s$^{-1}$.
\label{alpcenb2}}
}
\end{figure}

\subsection{The HERCULES Spectrograph}

The High Efficiency and Resolution Canterbury University Large \'Echelle Spectrograph ({\scshape Hercules}) (Hearnshaw et al.~2002) 
has been in operation at MJUO since 2001.  
It is fiber fed from the 1-m McLellan telescope, and is enclosed inside a vacuum tank ($P<0.005$\,atm).  
The spectrograph uses an R2 \'echelle grating in combination with a 200 mm collimated beam size.  
Cross dispersion uses a BK7 prism in double pass.  A large format CCD with 4130 x 4096 pixels each 15 $\mu$m in size allows complete wavelength 
coverage from 370 nm to 850 nm.  A 4.5 arcsec fiber feeding a 2.25 arcsec entrance slit gives a resolving power of $R = 70,000$ with high 
throughput ($\approx$ 12\% including telescope, fibers and CCD quantum efficiency).  The spectrograph is located in a temperature-stabilized 
room ($T\approx 20\pm0.05$\,C) and has no moving parts apart from focus which remains unchanged.  
An exposure meter is used to ensure precise control of exposure lengths and signal-to-noise ratios.
Because it was known that when observing bright stars the radial velocity precision of {\scshape Hercules} was limited by the effects of small 
guiding and centering errors, {\scshape Hercules} was also equipped with an iodine cell, which superimposes a dense reference spectrum on 
the stellar spectrum and also allows us to reconstruct the shape of the spectrograph's instrumental profile at the time of
observation. All RV results shown here were obtained with the iodine cell.

\subsection{HERCULES RV results for $\alpha$~Cen}

From 2007 to 2009 we collected RV data 
over 7 observing runs and 28 nights, with typical exposure times for component A ranging from 30 to 45 
s and for B from 90 to 120 s. The RV results have a long-term RV scatter of 2.5-3.0\,m\,s$^{-1}$ 
after subtracting the large trend due to the binary motion (see Figure~\ref{alpcen}). 
About 4\% of the data were rejected as outliers by a simple $\sigma$-clipping routine, removing all poor S/N-ratio spectra. 
In April 2009 we obtained more than 500 individual spectra for $\alpha$~Cen~B over the course of
5 nights. The weighted means of the nights have a very small rms scatter of 0.5\,m\,s$^{-1}$ (see Figure~\ref{alpcenb2}).
These first RV results demonstrate that we achieve a single shot RV precision of $\sim$2.5 \,m\,s$^{-1}$, sufficient to carry out this
program.  

\begin{figure}[t]
\includegraphics[angle=0,scale=.80]{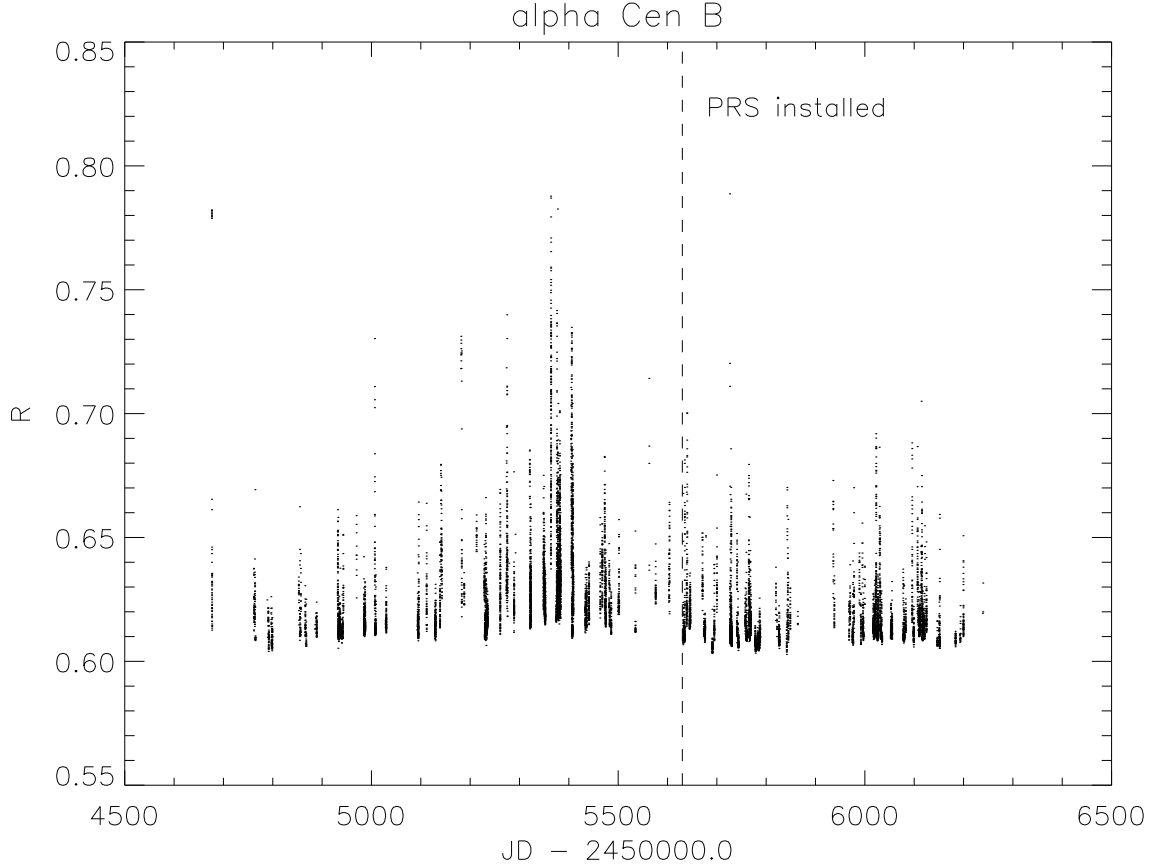}
\caption{Line index $R$ measurements using our $\alpha$~Cen~B spectra demonstrating the level of
spectral contamination from component A. A higher value of $R$ represents a higher level of contamination.
The installation of the Pinhole-Relay-System (PRS) reduced the amount of contamination somewhat, but
did not remove it. Note that without the PRS the amount of contamination would have steadily increased 
as the angular separation of the two stars is still shrinking.
\label{prs}}
\end{figure}

However, since 2010 the angular separation of the two stars has shrunk from $\approx$~7 to 5~arcsecs.
With a mean seeing at Mt\,John of 2 arcsec (with 95\% of the time better
than 3 arcsec) we have an increasing level of cross-contamination in our $\alpha$ Cen spectra.
The nominal 4.5 arcsec fiber of {\scshape Hercules} is simply too large for isolated observations of the two components of $\alpha$~Cen, as
discussed in Wittenmyer et al.~(2012). 

To minimise this effect we placed a pinhole at the 
focal plane of the telescope followed by a set of lenses to relay the light passing through the pinhole into the fiber. Apertures of 3 to 2~arcsec can be used to
reduce the size of the fiber on the sky. This pinhole relay system was installed in February 2011, and reduced, but
did not remove the spectral cross-contamination. Figure~\ref{prs} shows the level of contamination in our
data for $\alpha$~Cen~B (the effect for component B is, of course, stronger than for component A) as measured with a line index $R$. 
The quantity $R$ is defined as the ratio of the two line indices calculated as the flux in the line core as compared to the surrounding continuum for the 
H$\alpha$ and Na D lines, respectively. As the shapes of these
two strong lines are different for the A and B we can estimate the amount of contamination from the other star. 
The pinhole relay system also contains the iodine cell on a moving stage to insert it into the collimated beam between
the Cassegrain focus and the fiber entrance. It can be quickly removed for flat field spectra and Thorium Argon spectra, but inserted for the stellar spectra. 

\begin{figure}
\includegraphics[angle=0,scale=.80]{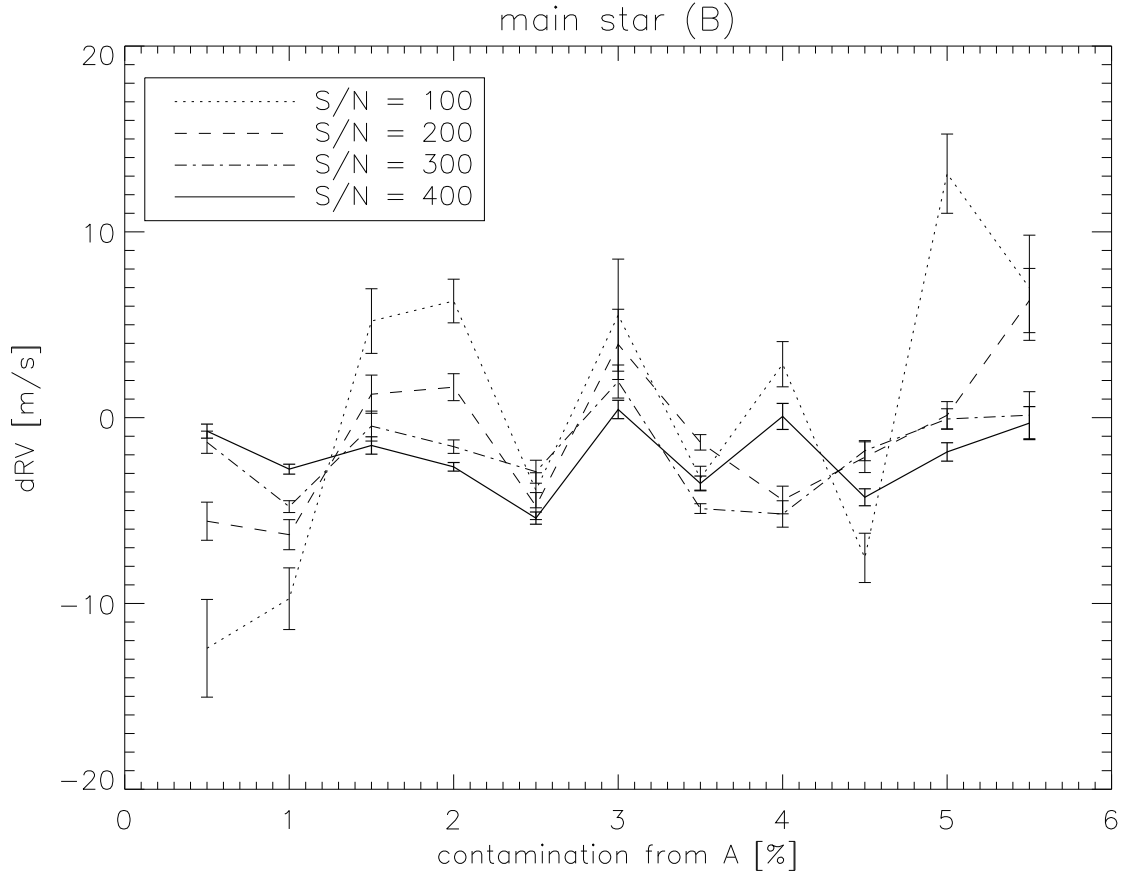}
\caption{RV results from tests performed with an advanced version of our {\sc Austral} iodine cell code that includes a second
low-level spectrum from a contaminating star. In this case we simulated that the target is $\alpha$~Cen~B with
varying amount of contamination from $\alpha$~Cen~A (x-axis gives amount of contamination in \%). The inital RV of the
uncontaminated spectra is at 0. 
We performed this test for 4 values of the S/N of the data. For each S/N value we created a set of 10 spectra. The
RV result is the mean value of these 10 spectra with the size of the error bar determined by the RMS scatter around this mean.
The RV separation for the two stars was set to 2.0\,km\,s$^{-1}$, which is the current RV separation of $\alpha$~Cen~A and B.
This shows that for S/N values in the 300-400 range, we can successfully recover the RV of the main target, nearly 
independent of the amount of spectral contamination.  
\label{chris1}}
\end{figure}

We pursue now an additional approach (Bergmann et al.~2012): to include and account for the spectrum of the contaminating star in our 
iodine cell data modeling pipeline {\sc Austral} (Endl et al.~2000). First test results for synthetic spectra with different amounts of contamination and 
signal-to-noise (S/N) levels are shown in Figure\,\ref{chris1}.     
These initial test results show that with a sufficiently high S/N level ($>300$) we are able to compute the RV of the main
target independent of the level of contamination from the second star. 
A detailed description of this new version of {\sc Austral} is the topic of a paper that is currently in preparation (Bergmann et al. in prep.). 
We use this new data modeling algorithm now to 
reprocess all of our 2010 to 2013 data for the $\alpha$~Cen system.

\section{Detectability of low-mass planets with the HERCULES program.}

We performed simulations to determine the expected detection efficiency of our program, once the effect of spectral contamination is removed.
We use the actual times of our current 19,316 observations of $\alpha$~Cen~B taken over 5 years and created artificial RV data sets with various different 
levels of white noise. The noise in our data is currently dominated by the effect of cross-contamination, which is highly systematic
in nature. This is a critical issue, as the strategy to detect low-amplitude signals with a large number of measurements
is only successful if the dominant noise in the data is random (white noise). Our goal is therefore to correct for the contamination and 
get as close as possible to white noise for the final noise structure in our RV data.     
The detection simulations will therefore inform our expected sensitivity to low-mass planets once the systematic noise from the
cross-contamination is removed.  

\begin{figure}[t]
\includegraphics[angle=270,scale=.60]{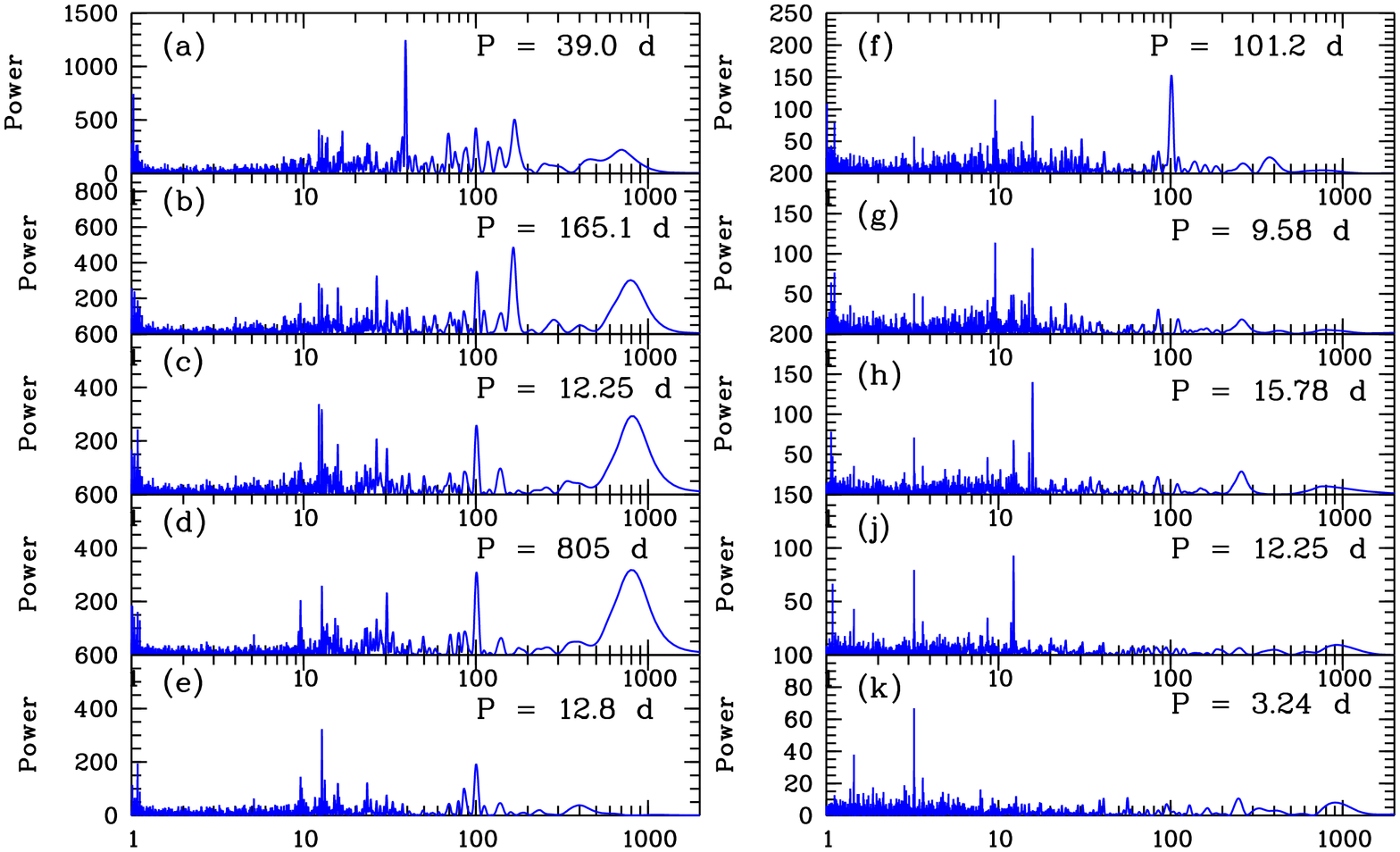}
{\caption{
Simulation of the RV detection of the purported Earth-mass planet in an 3.24-d orbit around $\alpha$~Cen~B (see text for details).
Panels (a) to (k) show the periodogram of the data after each step of pre-whitening. In each panel we state the period of the strongest 
signal that is subsequently fitted and subtracted from the data. Panel k displays the periodogram after removal of all
signals intrinsic to the star, the RV signal of the Dumusque et al. planet candidate with a period of 
3.24~d and an amplitude of $\pm0.40$\,m\,s$^{-1}$ (as reported in Hatzes 2013) is recovered with high confidence.  
\label{sim1}}
}
\end{figure}

\subsection{Sensitivity to the purported 3.24-d planet signal}

As a first test we simulated RV data that contains the purported 1.1 Earth-mass planet in a 3.24-d orbit as suggested by Dumusque et al.~(2012).
We created various white-noise RV data sets and injected the 8 signals intrinsic to the star (taken from Table~2 from Hatzes (2013)) with the
correct period, amplitude and phase. These 8 signals are the rotation period of $\alpha$~Cen~B of 39~d and its harmonics. We then added the 
planetary signal with a period of 3.24~d and an RV semi-amplitude of $0.40$\,m\,s$^{-1}$ and phase as reported in Hatzes~(2013).

We did not add the binary orbit to the simulated data as the effect of the second star is sufficiently linear over the the time of our 
observations. With the dense sampling of our data this slope is
very well defined and it makes no difference whether we create simulated RV data with a
slope, and then fit it and subtract it, or not. If our sampling would be
sparse, the fitting of the slope could affect longer periodic signals. But this is not the case, as we are
interested in signals with at least a few cycles covered by the survey combined with the very dense
sampling of the slope.
    
We then performed the same pre-whitening procedure as Dumusque et al.~(2012) and Hatzes~(2013): to identify the strongest peak in the Lomb-Scargle 
periodogram (Lomb~1976, Scargle~1982), then fit and subtract this signal and start over again. Figure~\ref{sim1} shows this step by step procedure 
for a simulation with a white noise level of 
$3.0$\,m\,s$^{-1}$. The final periodogram (panel k) shows very clearly that the 3.24-d signal is highly significantly detected by our survey at this noise level.
This is not surprising, as the {\scshape Hercules} survey covers about 600 cycles of this signal. 

\begin{figure}[t]
\includegraphics[angle=270,scale=.60]{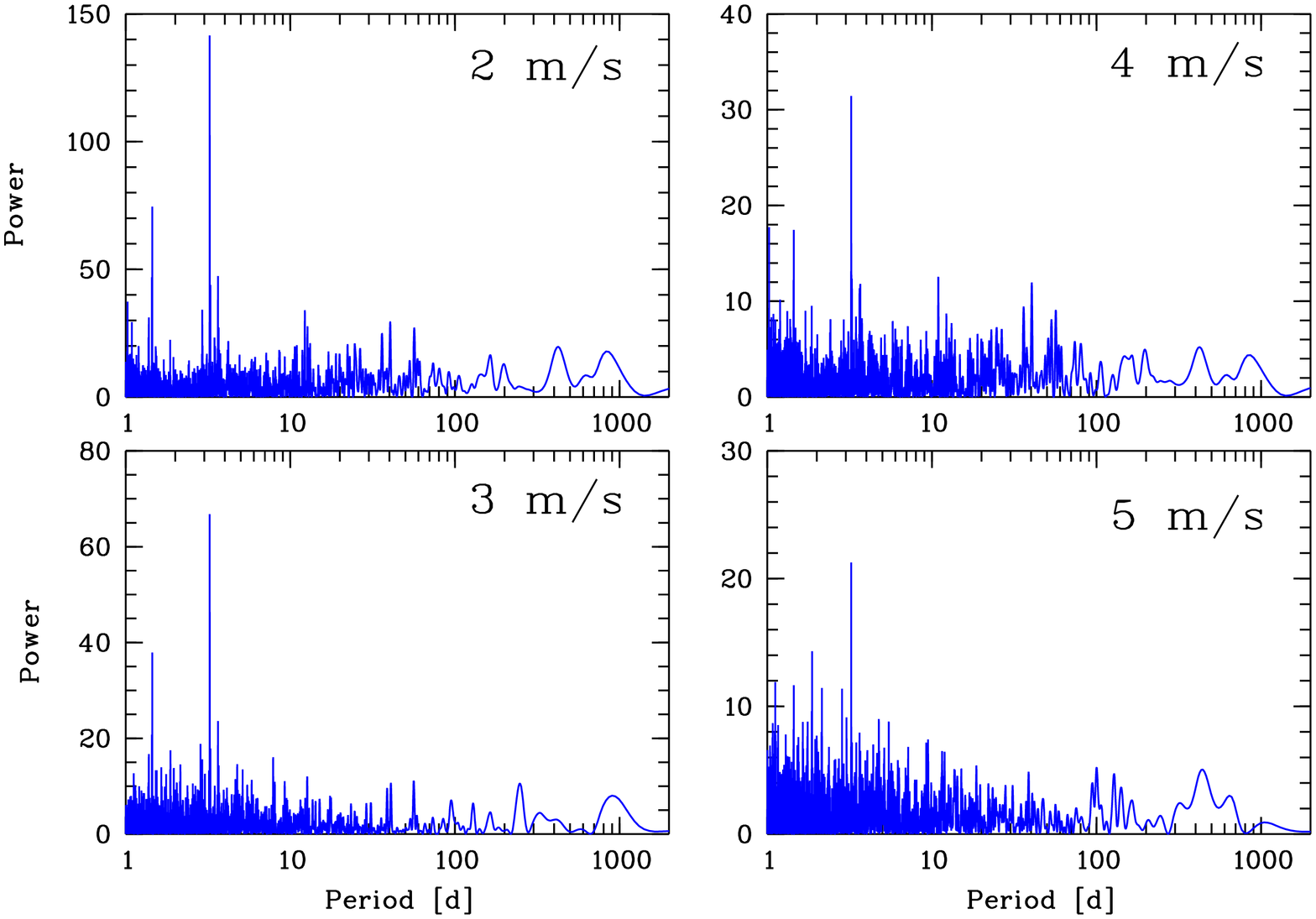}
{\caption{
Sensitivity to the 3.24-d signal as a function of white noise in our simulated RV data. For each case we performed the same 
pre-whitening procedure as shown in Figure\,\ref{sim1}. Each panel displays the final periodogram after subtraction of the
stellar intrinsic signals.   
\label{sim2}}
}
\end{figure}

Figure\,\ref{sim2} shows the results for four different simulations with white noise levels ranging from 2 to 5\,m\,s$^{-1}$. As expected, the 
significance of the signal of the low-mass planets drops with higher noise levels. Still, even with a relatively high noise level of
5\,m\,s$^{-1}$, we should be able to confirm this signal with the {\scshape Hercules} data, owing to the large number of measurements. 

\subsection{Sensitivity to planets in the habitable zone}

As the next test case we simulated RV data that contain the 8 stellar signals, as in the previous case, but no 3.24-d signal and instead 
injected
the RV signature of a super-Earth planet with a semi-amplitude of 0.35\,m\,s$^{-1}$ (minimum mass of 3.2~$M_{\oplus}$) and a period
of 234.3~d (at random phase). This period places the planet at the inner edge of the narrow habitable zone estimate given by 
Kaltenegger \& Haghighipour~(2013). As in the previous simulation we filtered the dominating stellar signals with a pre-whitening procedure until we
arrived at redsiduals that contain only noise and the planetary signal.  
   
\begin{figure}[t]
\includegraphics[angle=270,scale=.60]{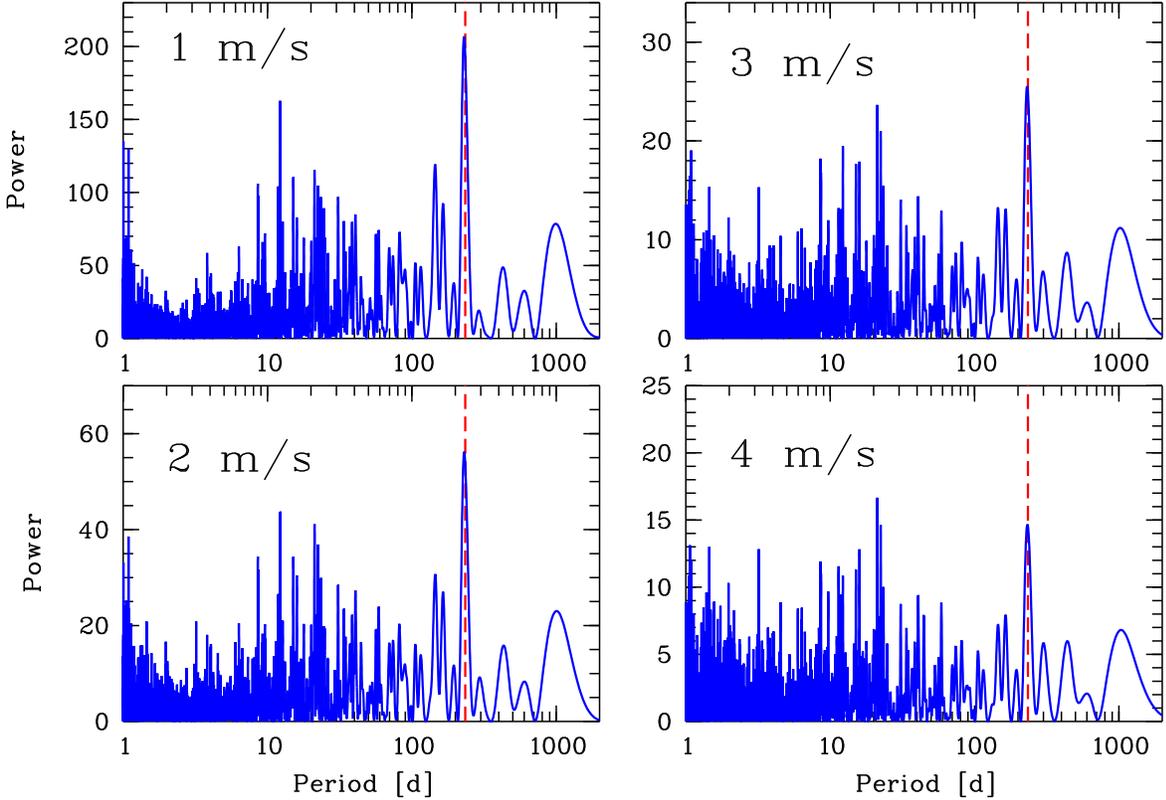}
\caption{
Period search results in our simulated RV data that includes a sinusoidal signal with 
$K=35$\,cm\,s$^{-1}$ with a period of 234 days (vertical dashed line) and after removal of the stellar intrinsic signals.  
Four different white noise levels (1 to 4\,m\,s$^{-1}$) are shown. The signal of the 3.2~$M_{\oplus}$ planet drops
below the noise between 3 and 4\,m\,s$^{-1}$.   
\label{sim3}}
\end{figure}

Figure~\ref{sim3} summarizes the results of four simulations using white noise from 1 to 4\,m\,s$^{-1}$.  
This shows that with a 2\,m\,s$^{-1}$ RV precision, we should be able to discover super-Earth planets in the habitable
zone of $\alpha$~Cen~B with the {\scshape Hercules} survey data. 

These results show us that we require our final white noise budget not to exceed the 3\,m\,s$^{-1}$ level. Based on our 
older data, we do expect final RV uncertainties to be below this level. This would mean that we are sensitive to 
super-Earths in the habitable zone of $\alpha$~Cen~B with the {\scshape Hercules} RV survey data.  
 
However, we used some assumptions in our simulations that might limit the significance of these results: (1) we used the
exact periods, phases and amplitudes of the signals intrinsic to the star as found by HARPS. We can expect that these signals
will show up differently in our {\scshape Hercules} data and (2) we only considered circular orbits. The periodogram is most
sensitive to sinusoidal variation and sensitivity to more eccentric orbits drops typically for $e>0.4$ (Endl et al.~2002). This
would mean that we could miss even more massive planets on eccentric orbits by using periodogram analysis alone.
We will therefore use various methods (Fourier-based pre-whitening, local de-trending, a genetic algorithm, etc.) 
to perform this signal extraction. A planetary-type RV signal must be successfully recovered by several techniques. 

\section{Conclusions}
The field of exoplanets has steadily moved forward to allow the detection of planets with masses similar to Earth and with 
orbital periods inside the circumstellar habitable zone of their host star. Owing to its proximity, the $\alpha$~Centauri system is a very
attractive target for such an intensive search for rocky planets using the RV technique. Any planets around the nearest stars
to the Sun would allow a large variety of follow-up investigations to study these planets for their potential
habitability. If the planet even happens to transit, we would be able to use JWST and the next generation
of extremely large telescopes to probe the atmospheres of such a planet using transmission spectroscopy.

Recently, Dumusque et al.~(2012)
announced the discovery of a very low-mass planet in a short 3.2~day orbit around $\alpha$~Cen~B using 4 years of HARPS data.
However, Hatzes (2013) casts doubt on the existence of this planet. Clearly, an independent falsification or confirmation
is needed. 

We are performing a concentrated observing campaign on the $\alpha$~Centauri system with the {\sc Hercules} spectrograph 
at the 1\,m McLellan telescope at Mt~John University Observatory in New Zealand. As of January 2014 we have observed over 26,000 spectra of $\alpha$~Cen~A and 
over 19,000 spectra of $\alpha$~Cen~B. The goal of our program is to achieve sensitivity to RV signals of rocky planets with 
orbital periods inside the circumstellar habitable zone.
Since 2010 we see the effect of cross-contamination in the spectra by the second star.
We have developed an advanced version of our RV code that can include and compensate for contamination in the data modeling to
compute the RV of the main target. The $45,000$ {\sc Hercules} $\alpha$~Cen spectra are currently in the process of being 
re-reduced with this new pipeline.

We explored the expected sensitivity of our program as a function of final noise level (after the removal of the systematic noise of
contamination). These simulations demonstrated that we should be able to confirm the purported Earth-mass planet with a period
of 3.24~d even with a high noise level of 5\,m\,s$^{-1}$. To be sensitive to super-Earths in the habitable zone we require the
noise budget to be below 3\,m\,s$^{-1}$. 

\acknowledgments
The Mt~John program is funded by Marsden grant UOC1007, administered by the Royal Society of New Zealand.
This project is also supported in part by the Australian Research Council 
Discovery Grant DP110101007. The authors would like to thank McDonald Observatory, University of Texas at Austin for 
allowing the use of the Sandiford iodine cell. SB acknowledges the support of McDonald Observatory which made the 2007 pilot study possible.
We thank the two referees, their comments were very helpful to improve this manuscript.

\end{document}